# Anisotropic magnetoresistance in topological insulator $Bi_{1.5}Sb_{0.5}Te_{1.8}Se_{1.2}$/CoFe heterostructures

B. Xia[1], P. Ren[1], Azat Sulaev[1], Z.P. Li[2], P. Liu[1], Z.L. Dong[2], and L. Wang[1*]

1. *School of Physical and Mathematical Science, Nanyang Technological University, Singapore, 637371*

*2. School of Materials Science and Engineering, Nanyang Technological University, Singapore, 638798*

Topological insulator is composed of an insulating bulk state and time reversal symmetry protected two-dimensional surface states. One of the characteristics of the surface states is the locking between electron momentum and spin orientation. Here, we report a novel in-plane anisotropic magnetoresistance in topological insulator $Bi_{1.5}Sb_{0.5}Te_{1.8}Se_{1.2}$/CoFe heterostructures. To explain the novel effect, we propose that the $Bi_{1.5}Sb_{0.5}Te_{1.8}Se_{1.2}$/CoFe heterostructure forms a spin-valve or Giant magnetoresistance device due to spin-momentum locking. The novel in-plane anisotropic magnetoresistance can be explained as a Giant magnetoresistance effect of the $Bi_{1.5}Sb_{0.5}Te_{1.8}Se_{1.2}$/CoFe heterostructures.

*Corresponding author: Lan Wang, Email: wanglan@ntu.edu.sg

## I. INTRODUCTION

Recently, a new type of topological condensed matter system characterized by $Z_2$ topological invariant, which is different from the Chern number topological invariant in quantum Hall system, has been proposed by theorists and then confirmed quickly by experimental physicists[1-5]. This new phase is called topological insulator. Topological insulator is composed of an insulating bulk state and time reversal symmetry protected two-dimensional spin helical surface states. Here, the spin-helicity means that the spin-up electrons propagate in one direction while spin-down electrons propagate in the opposite direction. The phenomenon of electron momentum determined spin orientation is called spin-momentum helical locking, which has been confirmed in recent angle-resolved photoemission (ARPES) experiments[6-13]. The protection against electron scattering was proved by scanning tunneling microscopy (STM) [14, 15]. Various novel electrical transport behaviors of topological insulators have been theoretically predicted and experimentally reported in recent literatures[16-37]. Surface transport induced Aharonov-Bohm oscillation was observed in $Bi_2Se_3$ and $Bi_2Te_3$ nanowires[16, 17]. Topological surface state induced Hall effect has been observed[23]. $\pi$ Berry phase and strong spin-orbit interaction induced weak antilocalization on the exotic topological surface state has been thoroughly studied both theoretically and experimentally [24-37]. It is well known that probing helical surface transport in electrical transport experiments is challenging due to the large bulk contribution in prototype topological insulator systems, such as $Bi_2Se_3$, $Bi_2Te_3$, *etc*. Recently, new topological insulators $Bi_2Te_2Se$[38, 39] and $Bi_{2-x}Sb_xTe_{3-y}Se_y$[40, 41, 52] with huge bulk resistivity (1 - 10 $\Omega$ cm) and dominated surface transport were synthesized. The topological surface states in the material systems have been confirmed by ARPES[41]. In our recent work, pronounced electrical field ambipolar effect has been realized in gated devices fabricated by 100 - 200 nm thick nanoflakes exfoliated from $Bi_{1.5}Sb_{0.5}Te_{1.8}Se_{1.2}$ crystals[52]. We also observed 2D weak antilocalization behavior in a 596 nm thick $Bi_{1.5}Sb_{0.5}Te_{1.8}Se_{1.2}$ single crystalline flake[52]. These results indicate that dominated surface transport can be realized in $Bi_{1.5}Sb_{0.5}Te_{1.8}Se_{1.2}$ single crystalline flakes with a thickness of several hundred nanometers. These materials offer a well characterized playground for studying the topological surface state.

Topological insulator/ferromagnet interface is a potentially very interesting system to investigate. Various novel devices based on ferromagnets and topological insulators have been proposed in recent years[42-47]. For example, spin and electron transport on topological insulators with ferromagnetic electrode has been studied theoretically[42]. Inverse spin-galvanic effect has been predicted to occur at the interface between a topological insulator and a ferromagnet[43]. The thin film heterostructure based on topological insulators and ferromagnet was believed to generate giant magneto-optical Kerr effect and universal Faraday effect[45]. However, as far as we know, the transport experiment on topological insulator/ferromagnet has never been attempted so far.

In this work, we measured the magnetoresistance of a $Bi_{1.5}Sb_{0.5}Te_{1.8}Se_{1.2}$/CoFe heterostructure. We observed unusual in-plane anisotropic magnetoresistance effect that is determined by the angle $\theta$ between the applied magnetic field and the current direction. With a model based on Giant magnetoresistance (GMR) effect due to the peculiar characteristic of the surface states of topological insulator, the spin-momentum locking, we successfully explain the novel in-plane anisotropic magnetoresistance. In this model, the two spin polarized electron transport channels are the spin polarized topological surface states and the ferromagnetic CoFe layer.

## II. EXPERIMENTS

High quality single crystals of $Bi_{1.5}Sb_{0.5}Te_{1.8}Se_{1.2}$ (BSTS) were grown by a modified Bridgeman method. High purity (99.9999%) Bi, Sb, Te, and Se with a molar ratio of 1.5:0.5:1.8:1.2 were first melted at 850 $^{o}$C in an evacuated quartz tube which was located vertically in a specially designed furnace with large temperature gradient. The temperature was then decreased to room temperature over 3 weeks. The characterization of structure and chemical homogeneity of these crystals are shown in Ref 52. The cleaved crystals were characterized by standard four contact electrical transport measurements. Ohmic contacts were made by room temperature curved silver paste. The magnetoresistance was measured on a horizontal rotator of a Physical Property Measurement System (PPMS, Quantum design). To study the exotic helical surface of the topological insulator with significant surface transport, 5 nm CoFe ($Co_{0.9}Fe_{0.1}$) and 3 nm $Al_2O_3$ protecting layer were sputtered on (001) surfaces of 50-100 μm thick single crystals at room temperature in a chamber with base pressure of 1 x $10^{-8}$ Torr. Standard four contact electrical transport measurements were employed to measure the magnetoresistance of the BSTS/CoFe heterostructures. Both the current and magnetic field were applied in the (001) plane. The BSTS single crystals and BSTS/CoFe interface were examined by high resolution transmission microscopy (HRTEM).

## III. RESULTS AND DISCUSSION

### A. The temperature dependence of resistivity

We have measured the R(T) curve of more than 15 pieces of 50-100 μm thick cleaved single crystals. All of them show similar R (T) behaviors. The resistance increases with decreasing temperature at high temperature regime, which indicates that the Fermi level is in the band gap. Only the R (T) curves of three samples are plotted in Fig. 1. We name the samples as S1, S2 and S3. The resistivity at low temperature of every sample is higher than 1 Ω·cm. Based on the recent electric transport results, the topological surface conductance contribution can get to ~10% in a ~ 100 μm thick BSTS sample at 10 K [40, 52]. As shown in the inset of Fig. 1, the temperature dependence of the magnetoresistance of the sample S3 changes from weak anti-localization type to normal parabolic type. The weak anti-localization type magnetoresistance is due to the π Berry phase of the helical surface state and the strong spin orbit interaction in the bulk[33, 34, 48]. The thickness dependence of weak anti-localization behavior indicate that the helical surface state contribute significantly to the weak anti-localization in a ~100 μm thick BSTS sample[52].

**B. In-plane anisotropic magnetoresistance measurements**

To study the exotic helical surface of the topological insulator with significant surface transport, we performed in-plane anisotropic magnetoresistance measurements on BSTS/CoFe (5 nm) heterostructures. For every resistance *vs.* magnetic field measurement, the angle θ between the current and applied magnetic field was fixed. A series of the measurements were performed by changing θ from $0^o$ to $360^o$ with $15^o$ interval. In the measurements, the samples were cooled down to 10 K with zero field. After that, the magnetic field was set to 10 kOe to saturate the ferromagnetic moment in CoFe layer, then swept to -10 kOe and finally drove back to 10 kOe. When the magnitude of the magnetic field was larger than 800 Oe, the data was collected every 1 kOe. Between -800 Oe and 800 Oe, the resistivity was measured every 10 Oe. Fig. 2(A)a - 2(A)p show sixteen magnetoresistance measurements of a BSTS/CoFe sample at 10 K at sixteen different θs, respectively. The BSTS crystal of this sample was cut from the same crystal of the sample S1 (Fig. 1). The current was applied to the direction B on the (001) plane as shown in diagram 2q below the Fig. 2(A) and Fig. 2(B). The measurement setup is shown in diagram 2r. When the magnitude of the magnetic field is larger than 800 Oe, the system shows positive magnetoresistance with no special characteristics at all θ values. Very unusual magnetoresistance behavior is observed in a field range between -800 Oe and 800 Oe. The two most striking characteristics are the disappearance of magnetoresistance at $0^o$ and $180^o$ (subtracting the background from the single crystal BSTS) and the square shape magnetoresistance shown at several θs ($30^o$, $120^o$, $150^o$, $210^o$, $300^o$ and $330^o$). The novel anisotropic magnetoresistance was confirmed in nine BSTS/CoFe samples. As the (001) plane has a $C_6$ symmetry, the current was applied along the A, B or C direction with $20^o$ interval as shown in the diagram 2q below Fig. 2(B). For each current direction, three samples were measured. S3a and S3b were exfoliated from the same crystal of sample S3, of which the R(T) curve is shown in Fig. 1. The magnetoresistance measurements of the BSTS(sample S3a)/CoFe heterostructure are shown in Fig 2(B), in which the current is applied to C direction. Its anisotropic magnetoresistance shows the same characteristic as that of the BSTS(sample S1)/CoFe (Fig. 2A). The current flow in BSTS (sample S3b)/CoFe is along direction A, which also shows similar anisotropic magnetoresistance behavior. From these measurements, we can obtain two important conclusions. Firstly, the novel square shape magnetoresistance can be confirmed in BSTS/CoFe structures fabricated from different BSTS crystals. Secondly, the current flowing along different crystalline direction shows negligible effect on the anisotropic magnetoresistance. We also studied the temperature dependence of the square shape magnetoresistance (sample S1) as shown in Fig. 3. All the measurements were carried out with $\theta = 30^o$, because the evolution of the square shape magnetoresistance is more easily to be studied. It is found that the square shape magnetoresistance disappears when the temperature is higher than 50 K. The high temperature magnetoresistance ($T > 50$ K) is composed of a positive magnetoresistance from BSTS single crystal and two very broad peaks (negative magnetoresistance) which may be originated from the CoFe layer. The square shape type anisotropic magnetoresistance and its temperature evolution are phenomena unforeseen in normal single layer ferromagnetic systems.

To study the exotic magnetoresistance behavior, we measured the anisotropic magnetoresistance of BSTS single crystal and 5 nm thick CoFe film on Si (111) (with the same crystal structure and similar lattice constant as BSTS (001) surface), respectively (Fig 4(A) and 4(B)). As shown in Fig. 4(B), the CoFe film on Si (111) shows standard anisotropic magnetoresistance of ferromagnetic material[49-51]. Its magnetoresistance value almost keeps constant for all θs although the shape of the magnetoresistance curve evolves with the variation of θ. This behavior is very different from that of CoFe/BSTS heterostructure, of which the magnetoresistance shows pronounced change with θ and is zero near $0^o$ and $180^o$ (subtracting the background from the weak anti-localization of the single crystal BSTS). The magnetoresistance of CoFe films on Si(111) also shows little change with increasing temperature due to the high Curie temperature. As shown in Fig. 4(A), the magnetoresistance of non-magnetic BSTS single crystal shows no special feature at all θs except the weak anti-localization behavior. Therefore, we can conclude that the unforeseen magnetoresistance of BSTS/CoFe is not the addition of the normal anisotropic magnetoresistance of BSTS single crystal and CoFe thin film. The peculiar characteristics of the novel anisotropic magnetoresistance of the BSTS/CoFe heterostructure can be summarized. Firstly, the magnetoresistance from ferromagnet disappears at certain angles ($0^o$ and $180^o$) and only the magnetoresistance from BSTS is shown. Secondly, square shape magnetoresistance is shown at certain angles. Thirdly, the novel magnetoresistance disappears when T > 50 K and the magnetoresistance of BSTS single crystal changes from weak anti-localization type to parabolic type in the same temperature range. All these characteristics are different from the magnetoresistance of CoFe thin film on Si (111), although the Si (111) has the same crystal structure and very similar lattice constant comparing with the (001) surface of BSTS single crystal.

**C. A model to explain the novel anisotropic magnetoresistance: spin-momentum locking induced Giant magnetoresistance**

As shown in the cross section TEM of BSTS/CoFe heterostructure (Fig. 5), the interface between BSTS single crystal and CoFe thin film is clear although not very sharp. From the STEM picture, we can conclude that the element distribution between BSTS and CoFe is pretty sharp, which indicates that the chemical reaction between BSTS and CoFe is negligible. As we grew the CoFe thin film at room temperature, this result can be expected. Therefore, the effect of an unknown compound formed at the interface can be ruled out.

In order to explain the novel anisotropic magnetoresistance, we propose that, due to spin-momentum locking, the BSTS/CoFe heterostructure in fact form a spin valve or Giant magnetoresistance (GMR) device. The novel anisotropic magnetoresistance is a GMR effect.

As shown in Fig. 6(a), a standard spin-valve or GMR device is composed of two ferromagnetic layers (FM) separated by a nonmagnetic metallic layer (NM). The ferromagnetic moment in one of the ferromagnetic layers is pinned by an adjacent antiferromagnetic layer (AFM), usually fixed during growth or cooling down process in a magnetic field (the pinned layer), while the ferromagnetic moment in the other layer can be flipped freely by magnetic field (the free layer). Based on Mott's two-current model, the resistance of the system is determined by the relative orientation of the ferromagnetic moments in the two layers. Low resistance is realized if the spin of the free layer is parallel to the spin of the pinned layer. When the spin of the free layer is switched to antiparallel configuration to the pinned layer, the spin-valve is at its high resistance state. For more general condition, the GMR effect is determined by cosδ, where δ is the angle between the spin orientations in pinned layer and free layer. The parallel condition of spin orientation means $\delta = 0^o$, while the antiparallel condition of spin orientation means $\delta = 180^o$. There will be no GMR effect if δ keeps as $90^o$ when we scan magnetic field. This condition can be realized when we sweep a magnetic field perpendicular to the fixed spin orientation in the pinned layer. Furthermore, if the magnetic field is along the easy axis of the free layer, almost all the spins in the layer will flip at the coercive field, therefore the resistance shows a sudden jump at the coercive field (square shape R (H) curve). On the other hand, if the magnetic field is not along the easy axis, the spin flipping is a process of gradual evolution, therefore the resistance of the spin valve structure varies slowly with changing magnetic field. The applied current in a GMR device is usually in-plane.

The key point here is that the BSTS/CoFe system forms a spin-valve or GMR system with one fixed spin-polarized layer and one free spin-polarized layer. As shown in Fig. 6(b), for the BSTS/CoFe heterostructure, there are one ferromagnetic layer and one adjacent helical surface layer. The ferromagnetic moment in the free CoFe ferromagnetic layer can be flipped by the applied magnetic field (the free layer), while in the top helical surface layer, the transporting electrons show fixed spin orientation (the fixed layer) due to the *spin-momentum helical locking*. In the presence of spin-momentum locking, the required conditions for realizing GMR effect, a “free” ferromagnetic layer and a layer with fixed spin orientation, are fulfilled in the BSTS/CoFe heterostructure system. One important point is that the topological surface state will not be destroyed by the magnetic moments in CoFe, because the ferromagnetic moments in CoFe are in-plane and therefore parallel to the BSTS surface. The thin oxide layer at the interface can also separate the CoFe layer and the BSTS surface states and therefore protect the spin-momentum locking. In this system, there are four electron transport channels, the CoFe ferromagnetic layer, the top helical surface contacting the CoFe, the bulk BSTS, and the bottom helical surface. Only the electrons scattered in the CoFe layer and the adjacent top helical surface contribute to the GMR effect.

Based on the above model, we can easily explain the aforementioned characteristics of the novel anisotropic magnetoresistance. The three characteristics of the anisotropic magnetoresistance CoFe/BSTS are the disappearance of anisotropic magnetoresistance at $0^o$ and $180^o$, the square shape magnetoresistance at $30^o$, $210^o$, $150^o$, and $330^o$, and the temperature evolution of the anisotropic magnetoresistance. As aforementioned, the GMR effect is determined by the angle between the spin orientation in the pinned layer and free layer. As shown in the schematic diagram (Fig. 6(b)), the spin orientation in helical surface is fixed and perpendicular to the current flowing direction. When $\theta = 0^o$ or $180^o$, the applied magnetic field (the spin orientation in the CoFe layer) is perpendicular to the fixed spin orientation in the helical surface state (the pinned layer), therefore there is no GMR effect except the background weak anti-localization effect from the bulk BSTS single crystal. With increasing θ (from $0^o$ to $90^o$ and from $180^o$ to $270^o$), the projection of the applied magnetic field on the fixed orientation of spins in the helical surface increases and therefore GMR effect gradually enhances. At θ = $90^o$ or $270^o$, the applied magnetic field and the fixed spin orientation are parallel or antiparallel to each other and therefore generate the largest GMR effect. If we further increase θ, the GMR will decrease again and get to zero at θ = $360^o$ and $180^o$. These phenomena are just what we observed in our measurements (Fig. 2). It should be noted that the GMR effect is only determined by the relative orientation of the current flowing and magnetic field. As the magnetic field is in-plane, the $180^o$ rotation of the sample is equivalent to the switching the current flowing direction to opposite direction, which will switch the spin orientation to opposite direction due to spin-momentum locking. As for the square shape magnetoresistance at $30^o$, $210^o$, $150^o$, and $330^o$, we believe that it is due to the magnetic anisotropy of the CoFe layer. As shown in Fig. 6(b), the BSTS/CoFe system has biaxial magnetic easy axes near θ = $30^o$ ($210^o$) and θ = $150^o$ ($330^o$). As discussed in the last paragraph, when the applied magnetic field is along the easy axis, all the spins in the free layer switches to the opposite direction simultaneously at the coercive field and therefore a square shape R (H) curve is obtained. The origin of the easy axis is still not clear yet. It may originate from the coupling between the helical surface state of $Bi_{1.5}Sb_{0.5}Te_{1.8}Se_{1.2}$ and CoFe, which diminishes at high temperature.

As the novel magnetoresistance behavior is due to the spin-momentum locking at the helical surface, it is also easy to explain the temperature dependence of the spin-valve type magnetoresistance as shown in Fig. 3. It should be noticed that the weak anti-localization behavior of BSTS (the inset of Fig. 1) and the GMR type magnetoresistance (Fig. 2) appear in the same temperature range and disappear with increasing temperature. The interconnected results indicate that the disappearance of GMR magnetoresistance with increasing temperature may be due to the loss of significant helical surface transport at high temperature. The high temperature magnetoresistance (T > 50 K) is composed of a magnetoresistance from BSTS single crystal and an anisotropic magnetoresistance from the CoFe layer. It also should be noted in Fig. 3 that the magnetoresistance shows two downward sharp peaks at 20 K. It might be due to the competing effect of thermal agitation energy and coupling between momentum-locked spin and magnetic moment of CoFe. It is well known that the GMR effect is only determined by the average spin polarization of electrons moving from high potential to low potential. A larger current density only means that the percentage of electrons moving to a certain direction becomes larger. The average spin polarization of the current will remain the same. As larger current density will not change the spin polarization of the current, the anisotropic magnetoresistance should remain the same with changing current density, which is also confirmed by our measurement as shown in Fig. 7.

## IV. CONLUSION

In summary, novel in-plane anisotropic magnetoresistance has been observed in topological insulator BSTS/CoFe heterostructure. To explain the novel effect, we propose that the BSTS/CoFe heterostructure forms a spin-valve or GMR device with one fixed spin polarized layer and one free spin polarized layer. The ferromagnetic moment in the free CoFe ferromagnetic layer can be flipped by the applied magnetic field (the free layer), while in the adjacent helical surface layer, the transporting electrons show fixed spin orientation (the fixed layer) due to the *spin-momentum helical locking*. As the magnetic moment of CoFe is parallel to the BSTS surface and the protection of the thin oxide layer, the CoFe cannot break the time reversal symmetry and therefore the spin-moment locking of the helical surface states is preserved. The novel in-plane anisotropic magnetoresistance is originated from the GMR structure formed by the CoFe layer and the adjacent helical surface of BSTS. Further theoretical and experimental research is required to fully understand the novel anisotropic magnetoresistance.

ACKNOWLEDGMENTS

Support for this wok came from Singapore National Research Foundation (RCA-08/018) and MOE Tier 2 (MOE2010-T2-2-059).

Figures

Fig. 1. Temperature dependence of the resistivity of three BSTS single crystals (S1, S2 and S3). The inset shows the temperature dependence of the magnetoresistance of sample S3 at 10 K, 30 K, 50 K and 100 K respectively.

Fig. 2. Anisotropic magnetoresistance curves of BSTS/CoFe heterostructures at sixteen different θs when current is applied to B direction (sample 1, Fig. 2(a)) and C direction (sample S3b, Fig 2(b)). The red curves are the scans from positive field to negative field. The black curves are the scans from negative field to positive field. The diagram 2q below the figure shows the A, B and C directions. The diagram 2r shows the set up of our experiments.

Fig. 3. The magnetoresistance evolution of BSTS/CoFe (sample S1) with increasing temperature. The range of magnetic field is from -800 Oe to 800 Oe. The magnetoresistance curves were measured at (a) 4 K, (b) 6 K, (c) 10 K, (d) 20 K, (e) 50 K, and (f) 100 K, respectively. θ is equal to $30^{o}$ for all measurements in this figure.

Fig. 4. (a) The anisotropic magnetoresistance of single crystalline BSTS with various θs at T = 10 K. (b) The anisotropic magnetoresistance of CoFe(5nm)/Si(111) with various θs at T = 10 K. The red curves are the scans from positive field to negative field. The black curves are the scans from negative field to positive field.

Fig. 5. (a) and (d) The high resolution cross section TEM of CoFe/BSTS. (b) The high resolution cross section STEM of the CoFe/BSTS. (c) The high resolution TEM of bulk BSTS single crystals.

Fig. 6. (a) The schematic diagrams of a normal Giant magnetoresistance device. The top FM layer is the free layer, while the bottom FM layer is pinned by the AFM layer. (b) The schematic diagrams of a Giant magnetoresistance device formed by a topological insulator/ferromagnet heterostructure. The FM layer is the free layer while the top surface helical state is the pinned layer. The spin orientation in the top surface helical state is perpendicular to the current flowing direction.

Fig. 7. Anisotropic magnetoresistance curves of BSTS/CoFe heterostructure (sample S3b, Fig. 1) at $\theta = 255^{o}$ with applied current of (a) 0.01 mA, (b) 0.025 mA (c) 0.05 mA (d) 0.1 mA (e) 0.15 mA and (f) 0.2 mA. The red curves are the scans from positive field to negative field. The black curves are the scans from negative field to positive field.

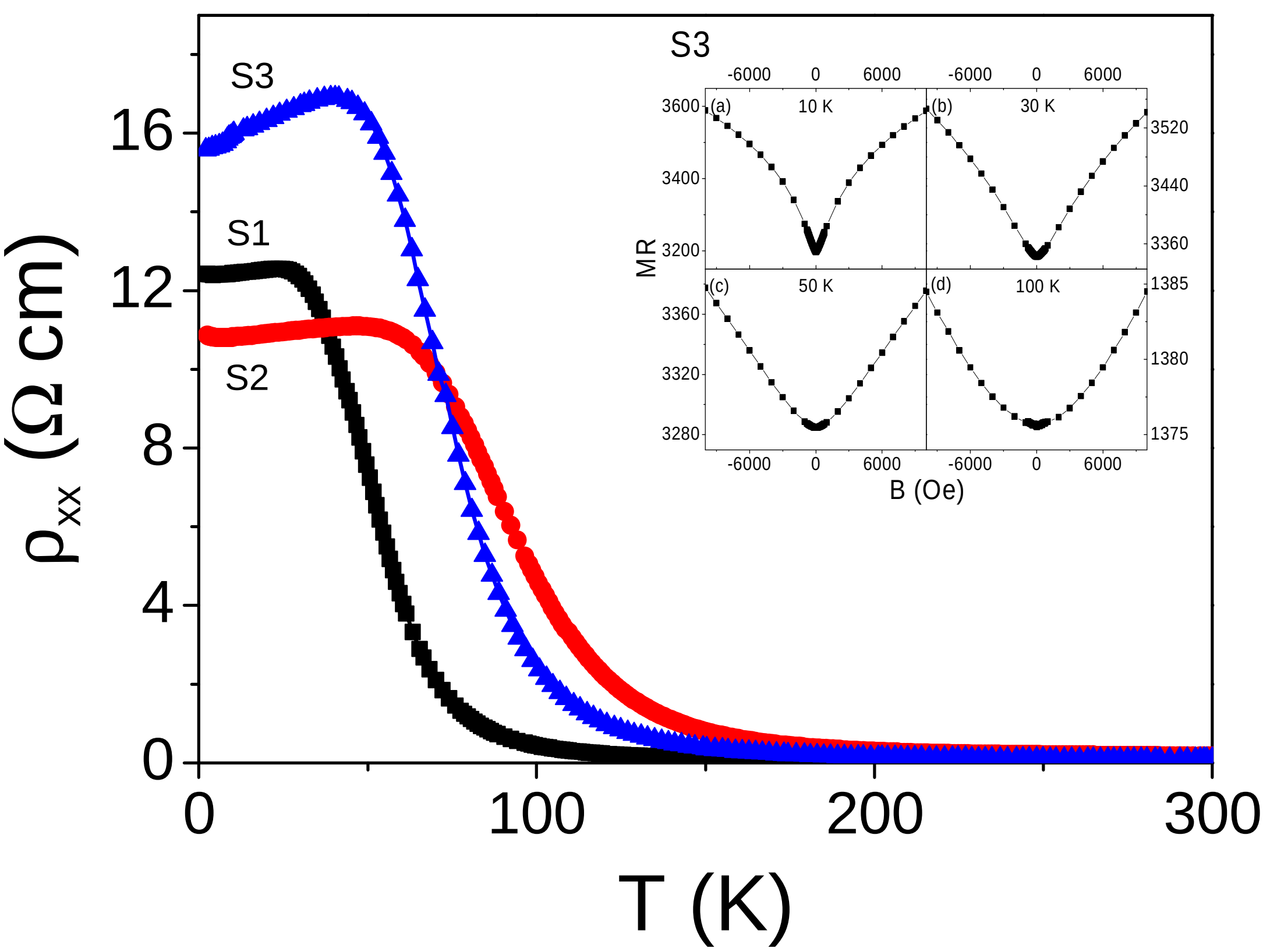

Fig. 1

(A)

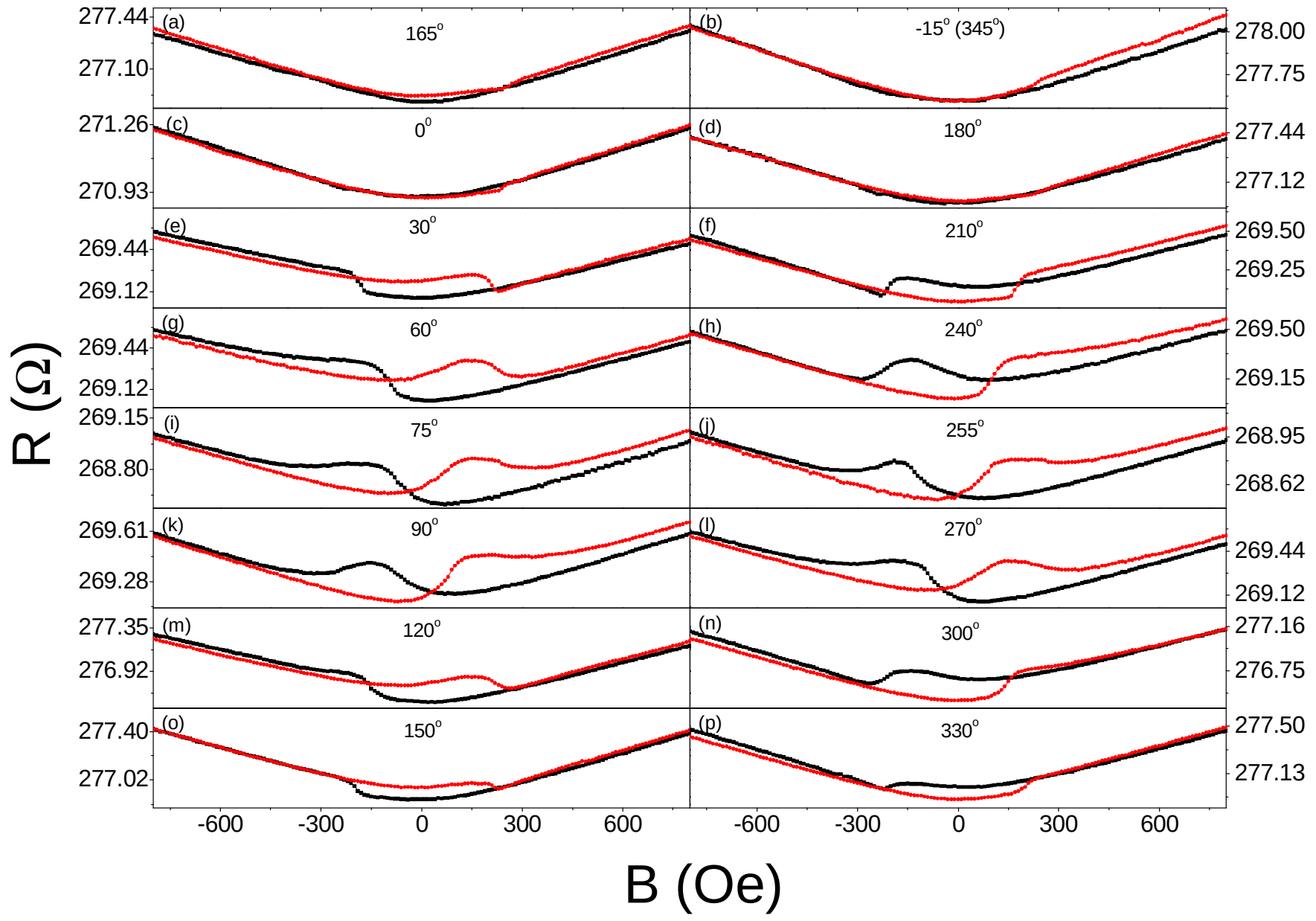

(B)

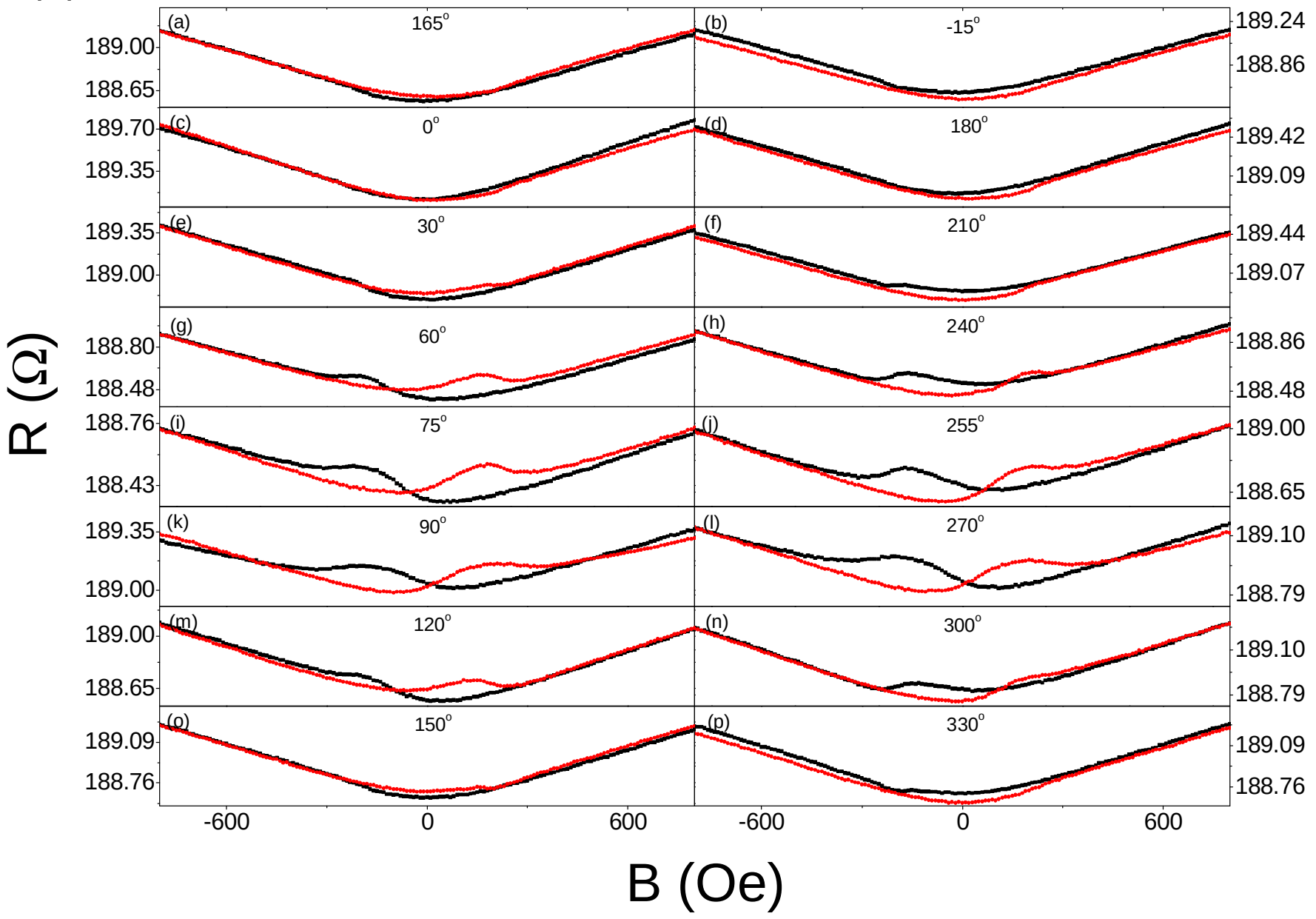

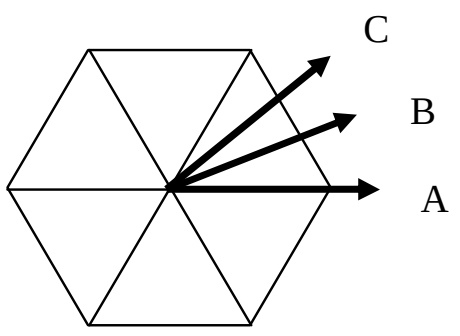

Diagram q

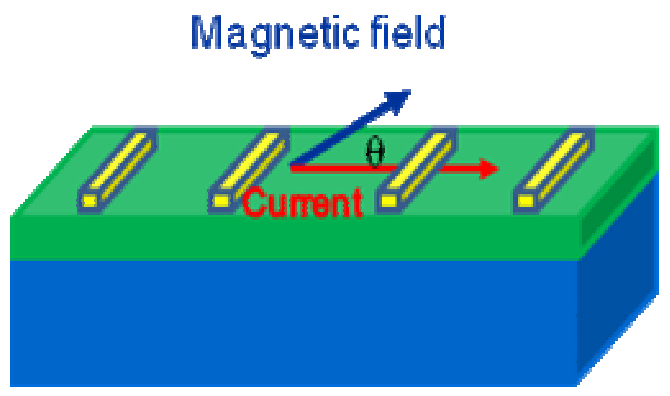

Diagram r

Fig. 2

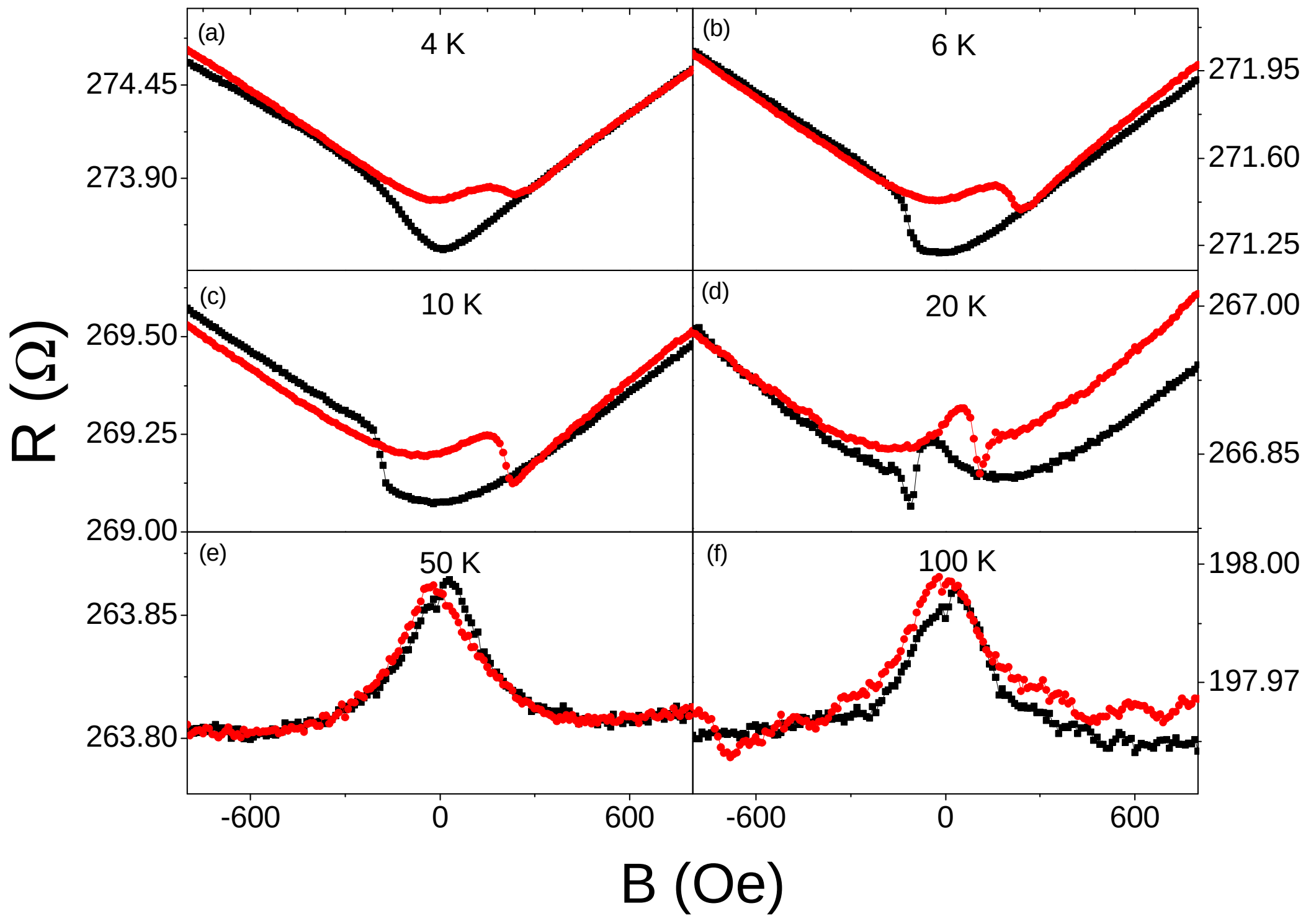

Fig. 3

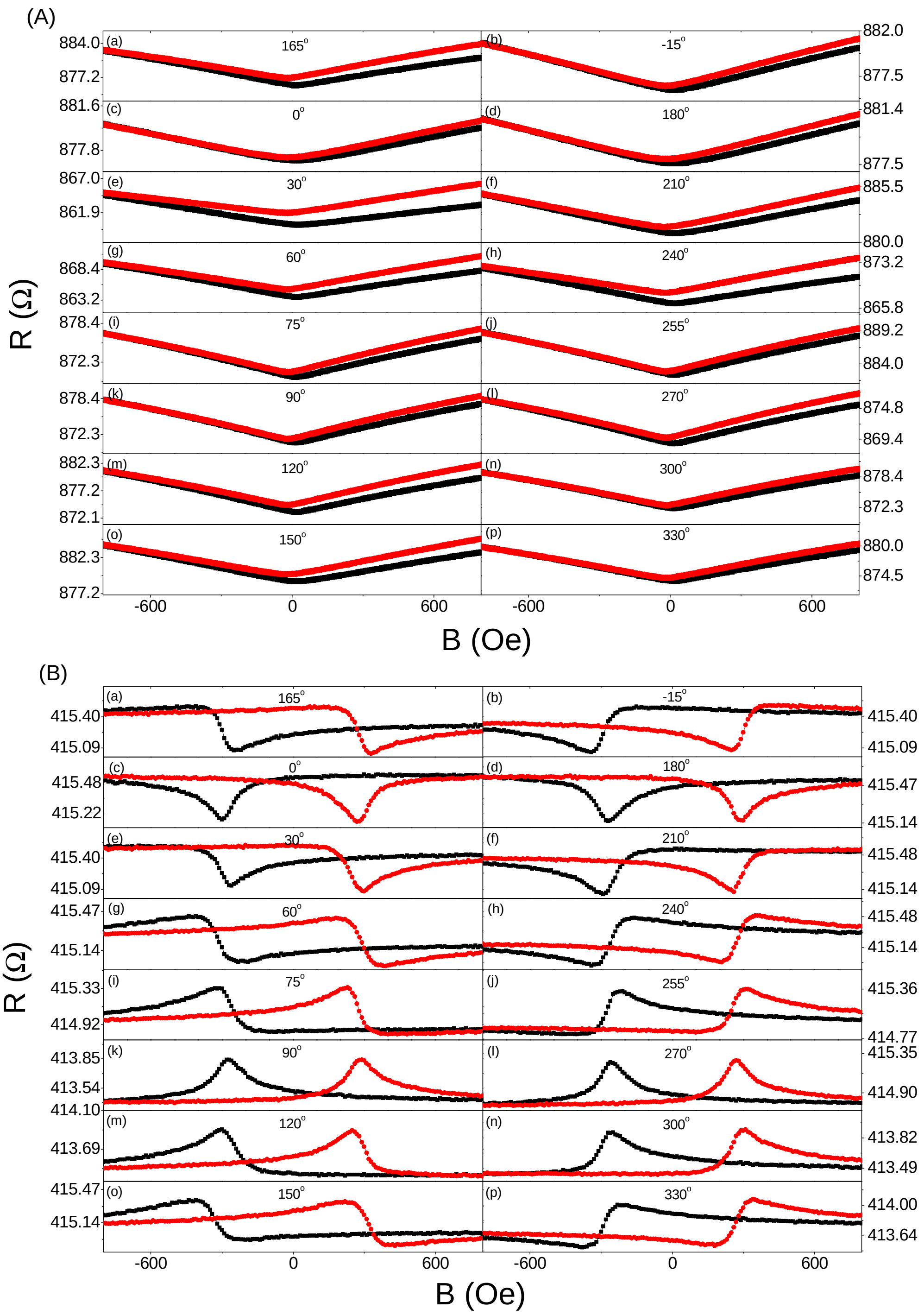

Fig. 4

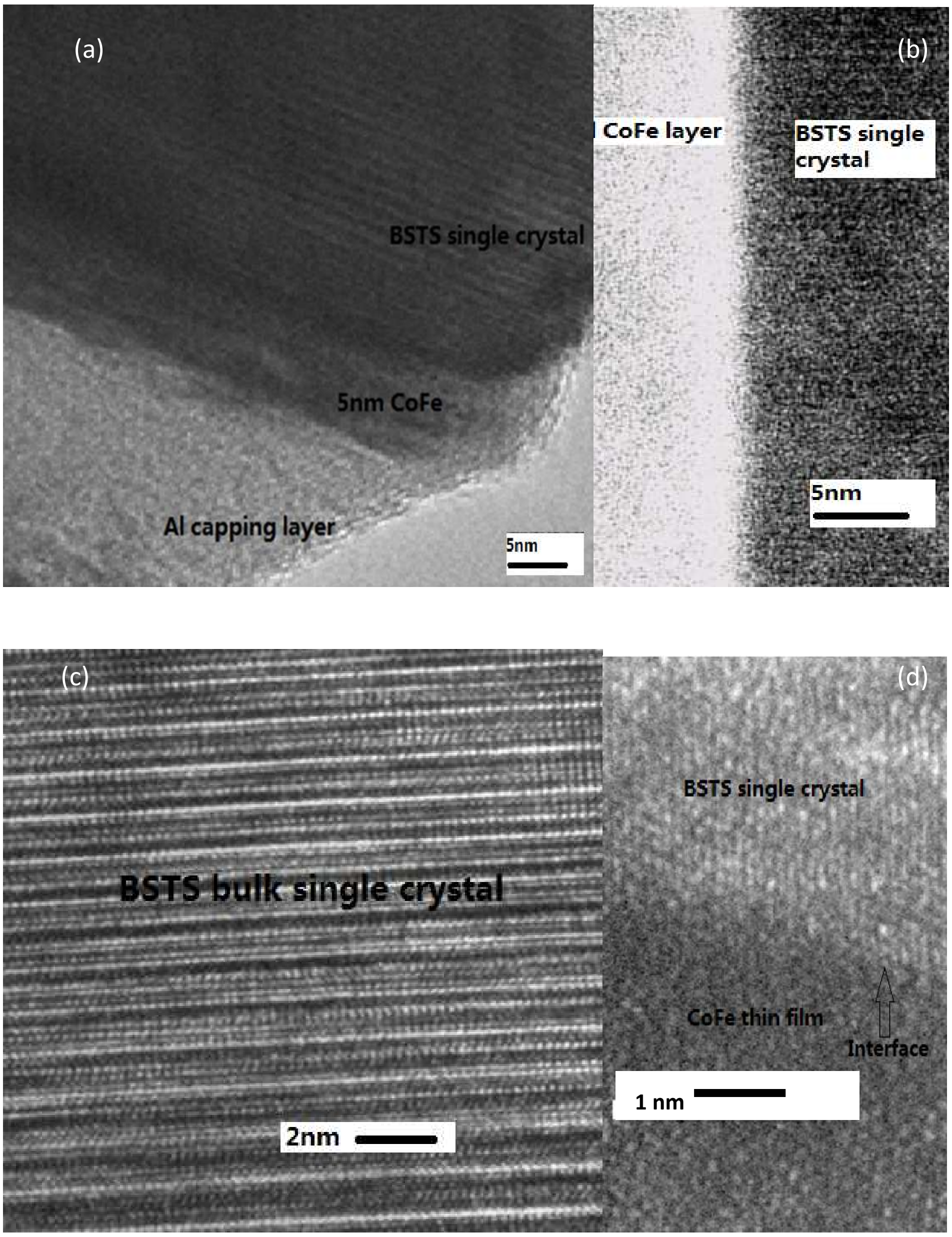

Fig. 5

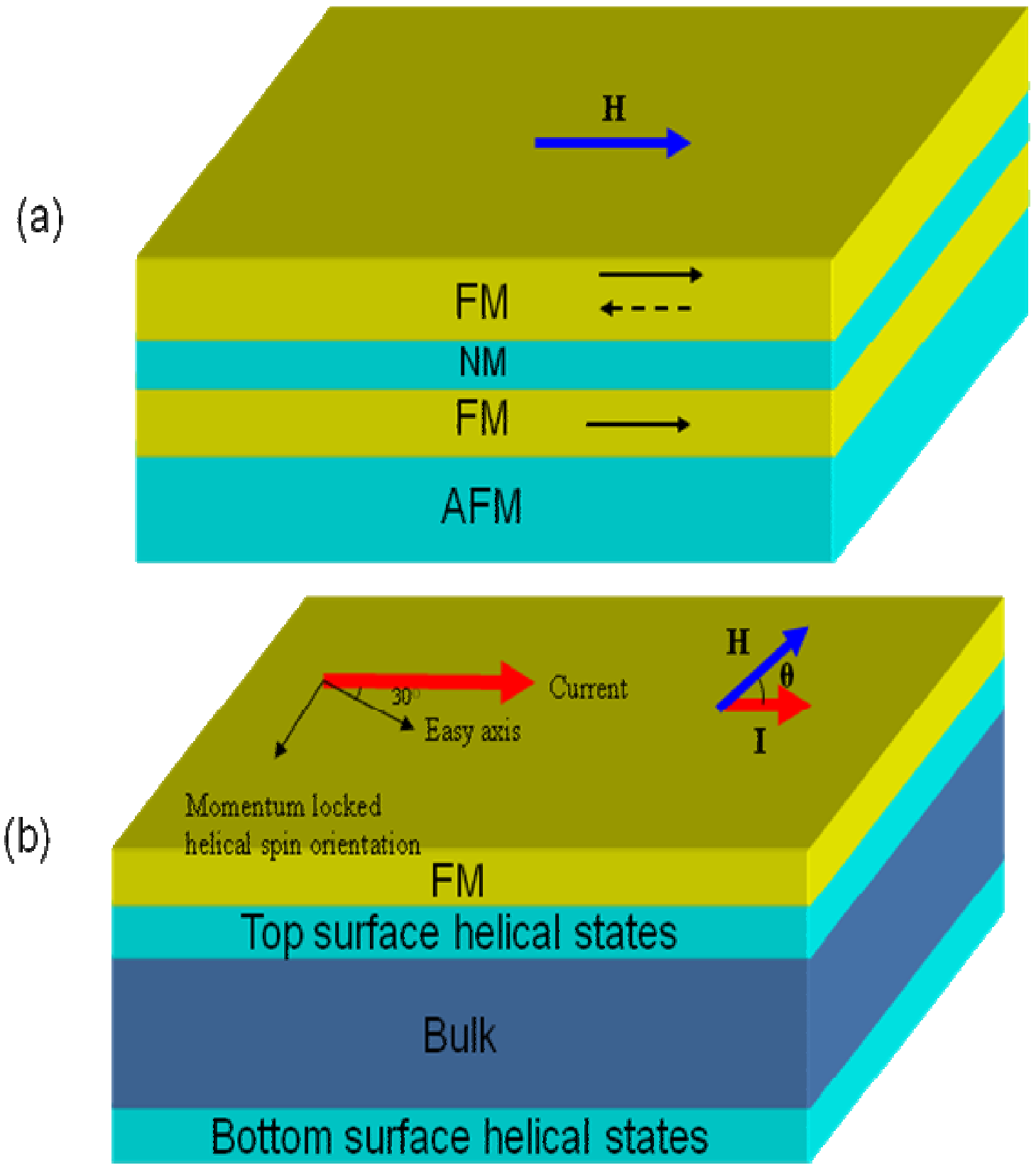

Fig. 6

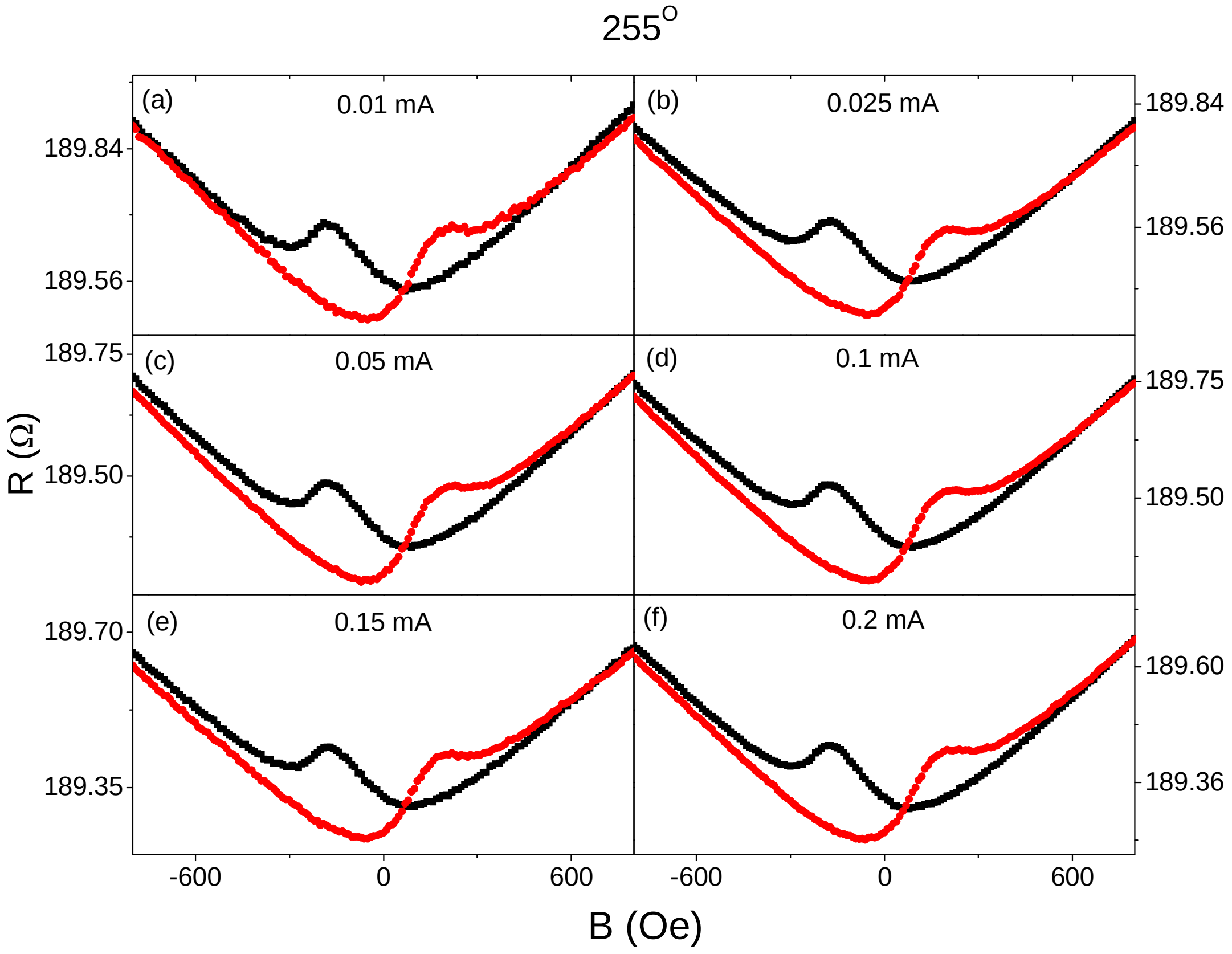

Fig. 7